\documentstyle[12pt,aaspp4]{article}

\newcommand{\lsun}{log $L/L_{\odot}\,$}
\newcommand{\msun}{$M/M_{\odot}\,$}

\begin{document}

\title{Improving the mass determination of Galactic Cepheids}

\author{G. Bono\altaffilmark{1}, W. P. Gieren\altaffilmark{2}, 
M. Marconi\altaffilmark{3}, P. Fouqu\'e\altaffilmark{4}, 
and F. Caputo\altaffilmark{5}}

\affil{1. Osservatorio Astronomico di Roma, Via Frascati 33,
00040 Monte Porzio Catone, Italy; Visiting Astronomer, 
ESO/Santiago, Chile; bono@mporzio.astro.it}
\affil{2. Dept. de Fisica, Grupo de Astronomia, Univ. de Concepcion, 
Casilla 160-C, Concepcion, Chile; Visiting Astronomer, ESO/Garching,
Germany; wgieren@coma.cfm.udec.cl}   
\affil{3. Osservatorio Astronomico di Capodimonte, Via Moiariello 16,
80131 Napoli, Italy; marcella@na.astro.it}
\affil{4. Observatoire de Paris-Meudon, DESPA F-92195 Meudon Cedex, France; 
and ESO, Casilla 19001, Santiago 19, Chile; pfouque@eso.org}
\affil{5. Osservatorio Astronomico di Roma, Via Frascati 33,
00040 Monte Porzio Catone, Italy; caputo@mporzio.astro.it}

\begin{abstract}
We have selected a sample of Galactic Cepheids for which accurate 
estimates of radii, distances, and photometric parameters are 
available. The comparison between their pulsation masses, based on 
new Period-Mass-Radius (PMR) relations, and their evolutionary masses,  
based on both optical and NIR Color-Magnitude (CM) diagrams, suggests  
that pulsation masses are on average of the order of 10\% smaller than 
the evolutionary masses. Current pulsation masses show, at fixed radius, 
a strongly reduced dispersion when compared with values published in 
literature. 
The increased precision in the pulsation masses is due to the fact that 
our predicted PMR relations based on nonlinear, convective Cepheid models 
present smaller standard deviations than PMR relations based on linear 
models. At the same time, the empirical radii of our Cepheid sample are 
typically accurate at the 5\% level.  

Our evolutionary mass determinations are based on stellar models 
constructed by neglecting the effect of mass-loss during the He burning 
phase. 
Therefore, the difference between pulsation and evolutionary masses 
could be intrinsic and does not necessarily imply a problem with either 
evolutionary and/or nonlinear pulsation models.  
The marginal evidence of a trend in the difference between evolutionary 
and pulsation masses when moving from short to long-period Cepheids 
is also briefly discussed. 
The main finding of our investigation is that the long-standing Cepheid 
mass discrepancy seems now resolved at the 10\% level either if account 
for canonical or mild convective core overshooting evolutionary models. 
\end{abstract}


\pagebreak 

\section{Introduction} 

Classical Cepheids are important objects not only because they 
are excellent distance indicators but also because it is possible, 
on the basis of both direct and indirect methods, to evaluate their 
intrinsic parameters such as radii, masses, and effective 
temperatures. A large amount of theoretical and empirical investigations 
has been devoted to classical Cepheids and the empirical uncertainties  
affecting their intrinsic parameters are by far smaller than for any 
other group of variable stars. As a consequence, Cepheids are key 
objects to constrain theoretical predictions.  

According to the so-called pulsation relation the period of a variable 
depends, at fixed chemical composition, on stellar mass, luminosity 
(radius), and effective temperature (Bono, Castellani, Marconi 2000a).
If we neglect the width in temperature of the instability strip the 
pulsation relation becomes a Period-Mass-Radius (PMR) relation.
This means that independent estimates of both period and mean radius 
supply an independent evaluation of Cepheid masses (called pulsation 
masses).  
Theoretical PMR relations based on nonlinear (Christy 1968; Fricke, 
Stobie, \& Strittmatter 1972, hereinafter FSS72) pulsation models 
have been widely adopted in the literature to estimate the Cepheid 
pulsation masses (Gieren 1989, hereinafter G89; Nordgren et al. 2000, 
hereinafter N00). 
On the other hand, Cepheid masses can be estimated on the basis of the 
Mass-Luminosity (ML) relation for intermediate-mass stars predicted 
by evolutionary models (Becker, Iben, \& Tuggle 1977; Stothers \& Chin 1994; 
Bono et al. 2000b; Baraffe \& Alibert 2001). This method requires 
a determination of the Cepheid distance and reddening as well as 
the use of a relation between period and bolometric correction (BC, 
as given in G89). A slightly different approach for deriving Cepheid 
evolutionary and pulsation masses has been recently devised by 
Beaulieu, Buchler, \& Kollath (2001, hereinafter BBK01). 
They derived, by adopting the Kurucz (1995) stellar atmosphere models, 
new analytical relations for both effective temperatures and BCs as 
a function of two observables, namely the period and the mean color 
(V-I).  
 
The comparison between pulsation and evolutionary masses led to  
the problem of Cepheid mass discrepancy (Cox 1980). Early estimates 
suggested that pulsation masses were approximately a factor of two 
smaller than the evolutionary masses. This problem has been 
substantially alleviated by the inclusion in the pulsation codes 
of the new OPAL (Rogers \& Iglesias  1992; Iglesias \& Rogers 1996) 
and OP (Seaton et al. 1994) radiative opacities (Moskalik et al. 1992; 
Kanbur \& Simon 1994). Despite the improvement in the accuracy of 
radiative opacities the agreement is far from being satisfactory, 
and indeed several doubts were raised for Magellanic Cepheids by 
Buchler et al. (1996) and by Wood, Arnold, \& Sebo (1997). 
Progress was made by Bono, Marconi \& Stellingwerf (2000c) for 
the group of Cepheids that show a bump along the light curve 
(Bump Cepheids). They obtained, on the basis of a fine grid of 
nonlinear, convective models relevant for Bump Cepheids  
($6.9 \le P \le 17.8$ days), a reasonable agreement between 
predicted and empirical masses. We have now to extend this 
work to the full period range covered by classical Cepheids, 
which is one of the main goals of this investigation. 
At the same time, we are also interested in checking the 
suggestion of BBK01, based on the OGLE database for Magellanic 
Cepheids, that current ML relations lead to evolutionary masses
that are systematically larger by $\approx0.1$ dex in \msun 
than the masses predicted by pulsation models.  

In \S 2 we discuss the theoretical framework and the selection 
of the Cepheid sample adopted in this investigation. In \S 3 we 
supply suitable analytical relations for fundamental (F) and first 
overtone (FO) pulsators connecting the stellar mass to periods, 
and radii. In this section we also estimate pulsation and evolutionary 
masses. Finally, in \S 4 we compare Cepheid masses based on pulsation 
and evolutionary predictions and provide plausible hypotheses to account 
for their difference.

\section{Theoretical models and empirical data}

Even though current spectroscopic measurements seem to suggest 
that the metallicity of Galactic Cepheids ranges from Z=0.008 to 
Z=0.024 (Fry \& Carney 1997), we adopted the chemical composition 
typical of solar 
neighborhood Cepheids, i.e. Y=0.28, Z=0.02. To supply homogeneous 
predictions for both F and FO PMR relations we used the same  
theoretical framework adopted by Bono, Caputo, \&  Marconi (1998),   
Bono, Marconi, \& Stellingwerf (1999, hereinafter BMS99) and by 
Bono et al. (2001, hereinafter BGMF01).   
To account for the current uncertainty in the predicted luminosity 
of intermediate-mass stars, we derived two PMR relations for 
canonical and noncanonical Cepheid models. These models were 
constructed by adopting two different ML relations based on evolutionary 
models that neglect (canonical, BMS99; Bono et al. 2000a) or 
include a mild convective core-overshooting (noncanonical, 
Girardi et al. 2000). 

The grid of canonical and noncanonical fundamental models 
constructed by BMS99 for stellar masses ranging from 5 to 11 $M_\odot$, 
were implemented with the new sequences of canonical models for 
\msun=4.5, 6.25, 6.5, 6.75 computed by BGMF01. We also constructed a 
new sequence of F models (\msun=4.0, \lsun=2.97) to properly cover the 
short-period range of noncanonical pulsators. 
As far as the first overtone is concerned, we adopted the 
canonical ($3.5\le$ \msun $\le5.5$) and the noncanonical 
($3.0\le$ \msun $\le4.75$) models provided by BGMF01.  
These models cover a large fraction of the Galactic Cepheid 
instability strip, and indeed the periods range from 2.7 to 
132.4 days for F and from 1.3 to 3.8 days for FO 
pulsators.  Table 1 lists the coefficients and the relative
errors of the analytical PMR relations for the two modes we 
studied. Note that these relations were derived by adopting 
nonlinear, convective models that show a stable limit cycle.  

To test the accuracy and plausibility of current models we also derived the
PMR relations for fundamental pulsators by adopting as independent variables
the stellar mass and the radius. We found:

\[
\log P (C) =-1.70(\pm0.01)-0.90(\pm0.12)\log M + 1.86(\pm0.04)\log R  \;\;\;\;\;\;\sigma=0.01
\]
\[
\log P (NC)=-1.74(\pm0.02)-0.70(\pm0.15)\log M + 1.80(\pm0.06)\log R  \;\;\;\;\;\;\sigma=0.01
\]

where the symbols have their usual meaning. These relations suggest that the
periods of fundamental Cepheids do not scale according to the period-density
relation  i.e. $P \propto R^{1.5}/M^{0.5}$, but as $P \propto R^{1.8}/M^{0.8}$.
However, the fundamental period of convective oscillating stellar envelopes
is proportional to $R^2/M$ as originally demonstrated by Gough, Ostriker, \&
Stobie (1965) on the basis of polytropic models.
The coefficients of current fundamental PMR relations are in good agreement
with these leading physical arguments. Note that the difference between the
periods predicted by the period-density relation and by nonlinear PMR relations
is vanishing for periods shorter than 10 days but becomes of the order of
25-30\% for periods equal to 60 days. 

In order to provide a detailed comparison between theory and 
observations  we selected the sample of radii, and distances 
determined by Gieren, Fouqu\'e, \& Gomez (1997,1998, hereinafter 
GFG97 and GFG98) for 34 Galactic Cepheids. The reason why we 
selected this particular sample is twofold:
i) radii and distances were derived by adopting for the entire sample 
the IR Barnes-Evans surface brightness technique (Fouqu\'e \& 
Gieren 1997). This approach presents several indisputable advantages 
when compared with other methods available in the literature and 
supply homogeneous and accurate evaluations of radii and distances
(GFG97).  
ii) for these objects both optical and NIR absolute magnitudes are 
available, as well as individual reddening estimates.  
We excluded from this sample the three longest period Cepheids 
(SV Vul, GY Sge, and S Vul) for the reasons already outlined by 
GFG98, and CS Vel since for this object no I band magnitude is 
available. However we included Polaris, since for this object an 
independent good estimate of the radius is available (N00). 
As a result, we end up with a sample of 31 Cepheids, which form 
the database for our current comparison between theory and 
observations.

\section{Pulsation and evolutionary masses}

Fig. 1 shows the pulsation masses for the Galactic Cepheids in our sample 
as a function of the radius.
The new masses were estimated by adopting the PMR relations listed in Table 1
and present a smaller dispersion when compared with similar estimates 
available in the literature (e.g. G89). In particular, we found that 
pulsation masses estimated by adopting the PMR relations derived by FSS72 
and Li \& Huang (1990) attain similar values but the intrinsic 
dispersion is at least a factor of two larger when compared with our 
estimates. The reason for this improvement is threefold:
i) previous relations rely on old input physics (opacities, and equation of 
state) and cover a narrower mass range; ii) the previous PMR relations 
are based on nonlinear, radiative models (FSS72), and linear, nonadiabatic 
models (Li \& Huang 1990), therefore they do not supply reliable estimates 
of the location of the red edge; iii) the small dispersion of pulsation 
masses is also due to the dramatic improvement in the accuracy of empirical 
radius measurements (the error bars are quite often smaller than the size 
of the symbols), when compared with former estimates (Cogan 1978; G89).  

As expected the Cepheid masses predicted by noncanonical PMR relations 
(triangles) are systematically smaller when compared with the canonical 
ones (filled circles). In fact, the He core masses of noncanonical models 
are, at fixed luminosity and chemical composition, larger than for 
canonical models. Data plotted in this figure present mild evidence 
that such a difference increases toward shorter periods. 
This effect could be due to the stronger sensitivity of the fundamental 
instability strip edges to stellar mass toward lower luminosities 
(Bono, Caputo, \& Marconi 2001).  
According to a recent detailed analysis of short-period Cepheids, 
Polaris (square) and SZ Tau (diamond) are almost certainly first 
overtone pulsators (N00; BGMF01). Unfortunately, the current 
uncertainties on the radius of SZ Tau does not allow us to constrain 
its pulsation mass. However, for Polaris we found that the pulsation 
mass is of the order of $M_p/M_\odot=4.9\pm0.1$. This error just accounts 
for the uncertainty due to the use of canonical and noncanonical 
PMR relations. This estimate is in very good agreement with the 
pulsation mass obtained by N00, i.e. $M_p/M_\odot=4.5\pm2$. 
The large difference in the error is due to the fact 
that we adopted PMR relations for FO pulsators, whereas the former authors 
were forced to fundamentalize the period of Polaris and then to use 
PMR relations for fundamental pulsators. In fact, up to now a PMR 
relation for FO Cepheids was not available. Note that the intrinsic 
dispersion of FO PMR relations is almost a factor of three smaller 
than for F pulsators (see last column in Table 1). 
As already noted for the Period-Radius relation (BGMF01) this difference 
is due to a decrease in the width in temperature of the instability 
strip and suggests that FO masses are practically unaffected by the 
intrinsic spread of the instability strip.  

In order to estimate the evolutionary masses we devised the following 
approach. We first plotted our Cepheid sample in the Color-Magnitude  
diagram (CMD), since it has been suggested that current evolutionary 
tracks are at odds with Magellanic short-period Cepheids (BBK01).
To account for subtle errors both in the distance, as well as in the 
reddening correction we adopted 
an optical ($M_V$, V-I) and a NIR ($M_K$, J-K) CMD (see Fig. 2).  
Evolutionary tracks for intermediate-mass stars ($3 \le$ \msun $\le 14$) 
recently computed by Bono et al. (2000b) for Y=0.29\footnote{Note that 
the marginal difference in the He content adopted for constructing 
the pulsation models (0.28 against 0.29) has negligible effects 
on current conclusions.} and Z=0.02 were adopted to construct 
isochrones ranging from 10 to 200 Myr. 
Theoretical predictions were transformed into 
the observational plane by adopting the bolometric corrections 
and the color-temperature relations by Bessell, Castelli, \& 
Plez (1998). A glance at the data 
plotted in the two panels shows that the empirical data are in very good 
agreement with stellar isochrones (see labeled ages). In particular, 
we find that predicted blue loops account for the distribution of 
observed Cepheids even at short periods.  

The evolutionary masses were estimated by adopting a detailed set of stellar
isochrones, with an age step of 2 Myr from 11 to 35 Myr, of 5 Myr from 35 
to 100 Myr, and of 10 Myr from 100 to 150 Myr. For each object the mass was 
estimated by performing an average over the masses that fall inside 
the empirical error box, i.e. the box given by the errors on distance 
and colors.  Fig. 3 shows evolutionary mass estimates based on both 
the (V, V-I) and the (K, J-K) CMDs as a function of period. 
The most important conclusion from this figure is that Cepheid 
masses based on optical and NIR CMDs agree at the level of few percent 
and are consistent within their uncertainties. However, evolutionary 
masses based on optical magnitudes seem to be slightly systematically 
larger than their NIR counterparts. This could be due to an error in 
the zero-point of the adopted reddening scale, since NIR magnitudes 
are practically unaffected by reddening corrections in contrast with 
optical magnitudes. The star-to-star differences between Cepheid masses 
based on optical and NIR magnitudes could be due to variable individual 
errors in the reddening correction to the V magnitudes. 
Moreover, it is worth mentioning that K magnitudes will be only marginally 
affected by the presence of an unresolved companion 
(physical or photometric blend), since Cepheid companions are expected 
to be on average hotter objects. This is not true for the V magnitudes
that can be substantially affected by the luminosity of the blue 
companion. This effect could also contribute to the variable mass 
difference for individual Cepheids which is observed in Fig. 3.

\section{Discussion and conclusions}

In order to perform a quantitative comparison between evolutionary and 
pulsational masses we performed a linear fit through current optical and 
NIR evolutionary masses and we found:

\[ 
\log M_e/M_\odot = -0.03(\pm0.02) + 0.48(\pm0.01)\log R/R_\odot\;\;\;\;\;\;\;\;\;\;\sigma=0.02
\]

where $\sigma$ is the standard deviation. The same fit performed over 
the pulsation masses based on the canonical ML relation yields: 

\[ 
\log M_p/M_\odot = -0.09(\pm0.03) + 0.48(\pm0.03)\log R/R_\odot\;\;\;\;\;\;\;\;\;\;\sigma=0.03.
\]

The two fits yield, within the small error, identical slopes; 
however, they show that evolutionary masses are systematically 
larger when compared with the pulsation ones. The difference ranges  
from $\approx$0.05 dex at $\log R/R_\odot$=1.6  to $\approx$0.04 at 
$\log R/R_\odot$=2.2. Taken at face value this difference implies a 
discrepancy between evolutionary and pulsation masses of the order 
of 13\% for short-period Cepheids and of the order of 10\% for 
long-period ones. Within current empirical uncertainties we cannot 
assess whether this trend is real or caused by the small number 
of long-period Cepheids included in our sample. The same result 
is demonstrated in yet another way i.e. by plotting the Cepheid 
masses as function of period. 
Fig. 4 shows the ratio between pulsation and evolutionary masses 
versus period. The pulsation masses are the canonical ones, 
while the evolutionary masses are a mean between the mass 
evaluations based on optical and NIR magnitudes. 
The pulsation masses are systematically smaller than 
the evolutionary masses by $\approx10$\% and the spread is of 
the order of 10\%. Data plotted in Fig. 4 should be compared with 
Fig. 5 in G89 to appreciate the improvement in the accuracy of 
current mass determinations. 
The sample of Galactic Cepheids we are dealing with is too small 
to reach any firm conclusion on the occurrence of a trend with the 
pulsation period. 

However,  
in this context it is worth mentioning that evolutionary masses 
are based on stellar models that neglect mass-loss during He-burning 
phases, whereas the pulsation masses do estimate the actual mass 
of Cepheids. Therefore, the mass difference we obtain could be a real 
feature, reflecting a mass loss of Cepheids of the order of 10\% during 
the He burning phase, and not the consequence of errors in the  
predictions based on evolutionary and/or pulsation models.  
As our most important conclusion,  we stress that the long-standing 
discrepancy between 
evolutionary and pulsation masses has been brought down, for the 
first time, to the 10\% level. 
Moreover, we point out that the current small residual discrepancy 
between evolutionary and pulsation masses is not affected by the adopted 
evolutionary scenario. In fact, pulsation masses based on noncanonical 
PMR relations are smaller when compared with the canonical ones, but the 
evolutionary masses based on noncanonical evolutionary tracks are smaller 
as well. Thus the ratio between the two mass estimates does not change. 
The comparison with the G89 results discloses that we have gone a long 
way since then in improving Cepheid models and empirical determinations 
of Cepheid radii and distances. 

Regarding the possibility of mass loss we are not aware of any 
recent spectroscopic investigation aimed 
at measuring the efficiency of mass-loss among both Galactic and 
Magellanic Cepheids. Empirical estimates based on infrared (IRAS) 
and ultraviolet (IUE spectra) emissions for a sizable sample of 
Galactic Cepheids suggest mass loss rates ranging from  $10^{-10}$ 
to $10^{-7}$ $M_\odot\, yr^{-1}$ (Deasy 1988). However, observational 
determinations 
are not well constrained, and indeed VLA observations (Welch \& Duric 1988) 
and resonance absorption line profiles (Rodrigues \& B\"ohm-Vitense 1992) 
provide upper limits ranging from $10^{-10}$ to $10^{-7}$ 
$M_\odot\,yr^{-1}$. In spite of the empirical uncertainties these values 
seem to support our hypothesis that the discrepancy between evolutionary 
and pulsation masses could be due to mass loss. In fact, the He burning 
lifetimes for a 5 $M_\odot$ Cepheid at solar chemical composition is 
of the order of 20 Myr while for a 11 $M_\odot$ Cepheid it is 2.5 Myr. 
The lack of a firm empirical evidence concerning the correlation of the 
mass loss efficiency with the pulsation period limits the significance
of the result. Future theoretical and empirical investigations to shed 
new light on Cepheid mass loss will be of great value.   

Current results seem to suggest that the discrepancy between 
evolutionary and pulsation masses for Galactic Cepheids is roughly 
a factor of two smaller (0.05 against 0.1 dex) than found by BBK01 for 
Magellanic Cepheids. Since, we adopted a different approach to estimate 
both evolutionary and pulsation masses, we cannot 
assess whether this difference is due to an intrinsic feature of  
Galactic Cepheids or to unknown systematic errors. Note that 
our pulsation masses based on the PMR relations do rely on periods 
and mean radii, whereas the approach suggested by BBK01  
requires periods, mean magnitudes, colors, and reddening 
corrections. Therefore our approach seems to be more straightforward 
and less vulnerable to errors in the adopted empirical data. 
On the other hand, evolutionary masses were estimated 
by comparing theory and observations directly into the CMDs, 
whereas BBK01 adopted different ML relations.  
New empirical estimates of Magellanic Cepheid radii and distances 
based on the infrared Barnes-Evans surface brightness technique 
will be crucial to address in the near future this apparent 
discrepancy between our results and those of BBK01.  
 
In passing we note that the evolutionary mass of SZ Tau is in good 
agreement with the Cepheid masses of short-period Cepheids. This 
suggests that the distance modulus is not affected by large 
uncertainties while the empirical radius, and in turn the pulsation 
mass, are affected by larger errors. 
Even though the distance of Polaris has been recently
estimated by Hipparcos (Feast \& Catchpole 1997) we cannot 
estimate its evolutionary mass, since I, J, and K magnitudes 
are not available for this object.

\acknowledgments
It is a pleasure to thank S. Cassisi for sending us the detailed set 
of isochrones at solar chemical composition and for several thorough 
discussions on the mass loss efficiency among intermediate-mass stars. 
We also wish to thank an anonymous referee for his/her pertinent 
suggestions that improved the readability of the paper.  
GB \& MM acknowledge financial support by MURST-Cofin 2000, under the 
scientific project "Stellar observables of cosmological relevance".
WPG gratefully acknowledges partial financial support received from 
Fondecyt projects 1000330 and 8000002.
Part of this work was carried out while WPG was a scientific visitor 
at the European Southern Observatory in Garching. WPG is grateful for 
the support provided by ESO. He would also like to acknowledge 
computer support from the Center for International Migration (CIM) 
in Germany. 

\pagebreak

\pagebreak
\begin{center}
\begin{tabular}{ccccc}
\tablewidth{0pt}\\
\multicolumn{5}{c}{TABLE 1. PMR relations at solar chemical composition:}\\ 
\multicolumn{5}{c}{$\log M = \alpha\,+\, \beta\log P \,+\,\gamma\log R^a$}\\
\hline
                     & $\alpha$  & $\beta$ &  $\gamma$& $\sigma$ \\
\hline
\multicolumn{5}{c}{Fundamental} \\  
 C$^b$      & $-0.966\pm0.012$ & $-0.599\pm0.078$ & $1.291\pm0.118$ & 0.011\\  
 NC$^b$     & $-1.073\pm0.015$ & $-0.608\pm0.133$ & $1.334\pm0.200$ & 0.014\\  
\multicolumn{5}{c}{First Overtone} \\  
 C$^b$      & $-2.776\pm0.004$ & $-1.661\pm0.140$ & $2.682\pm0.185$ & 0.004\\  
 NC$^b$     & $-2.034\pm0.004$ & $-1.179\pm0.246$ & $2.068\pm0.334$ & 0.004\\  
\hline
\end{tabular}
\end{center}
\begin{minipage}{1.00\linewidth}
\noindent $^a$ Stellar masses and radii are in solar units, while the 
periods are in days. $^b$ PMR relations based on Cepheid models constructed 
by adopting canonical (C) and noncanonical (NC) ML relations.  
\end{minipage}

\pagebreak
\figcaption{Pulsation masses as a function of radius for our 
Cepheid sample (see text).  
Solid circles and triangles refer to mass estimates of fundamental mode  
variables based on canonical and noncanonical ML relations respectively. 
The square and the diamond mark two first overtones, namely Polaris and SZ Tau. 
The error bars on the mass account for the uncertainty in the coefficients
of the PMR relations. The error bars in the mean radius are typically 
smaller than the size of the symbol.}   

\figcaption{Top: Color-Magnitude diagram ($M_V$, $(V-I)_0$) for our sample 
of Galactic Cepheids. Solid lines show theoretical isochrones at solar 
chemical composition for stellar ages ranging from 18 to 130 Myr. 
The error bars account for the uncertainty both in the distance modulus 
and in the reddening correction. Bottom: same as the top panel but for 
NIR bands ($M_K$, $(J-K)_0$). The ages of individual isochrones are 
also labeled.}  

\figcaption{Evolutionary masses as a function of radius for our sample. 
Solid circles
and triangles refer to mass estimates based on the optical and NIR CMDs,  
respectively. The evolutionary masses were estimated by adopting stellar 
isochrones, and in turn evolutionary tracks that neglect mass loss 
($\eta=0$) and convective core overshooting (canonical). The solid
line shows the linear fit between empirical radii and Cepheid masses.}     

\figcaption{Ratio between pulsation and evolutionary masses versus 
logarithmic period for the fundamental Cepheids in our sample. 
Pulsation masses were estimated by adopting the canonical PMR relation,
while the evolutionary ones are the mean between the mass estimates 
based on optical and NIR magnitudes.} 
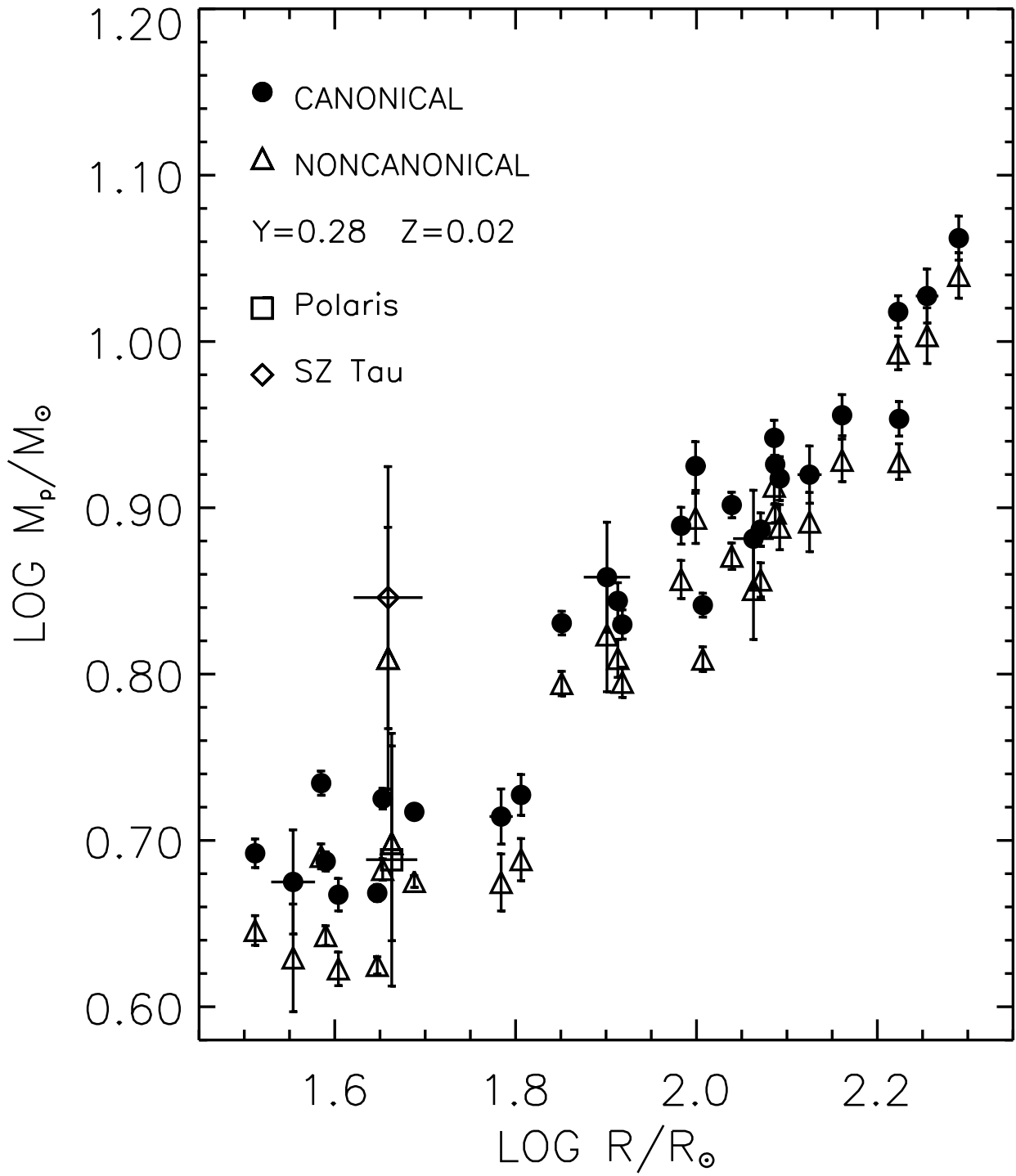
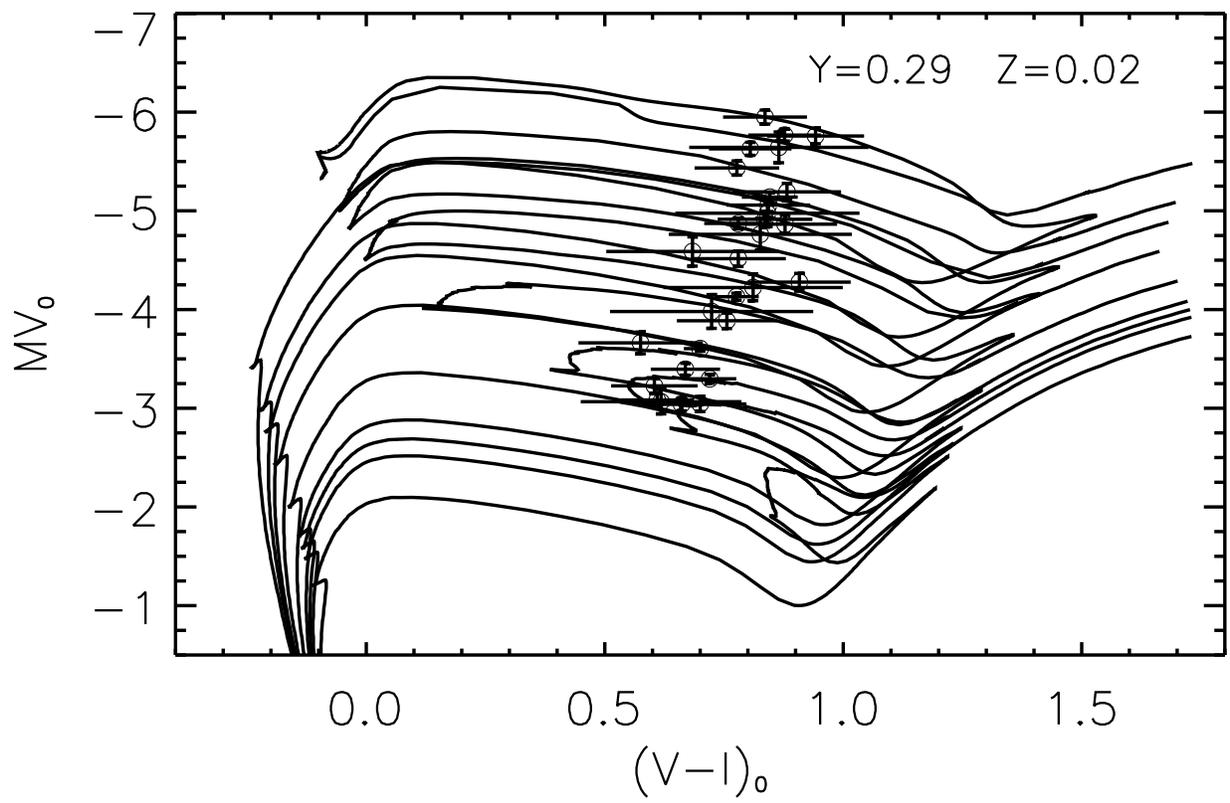
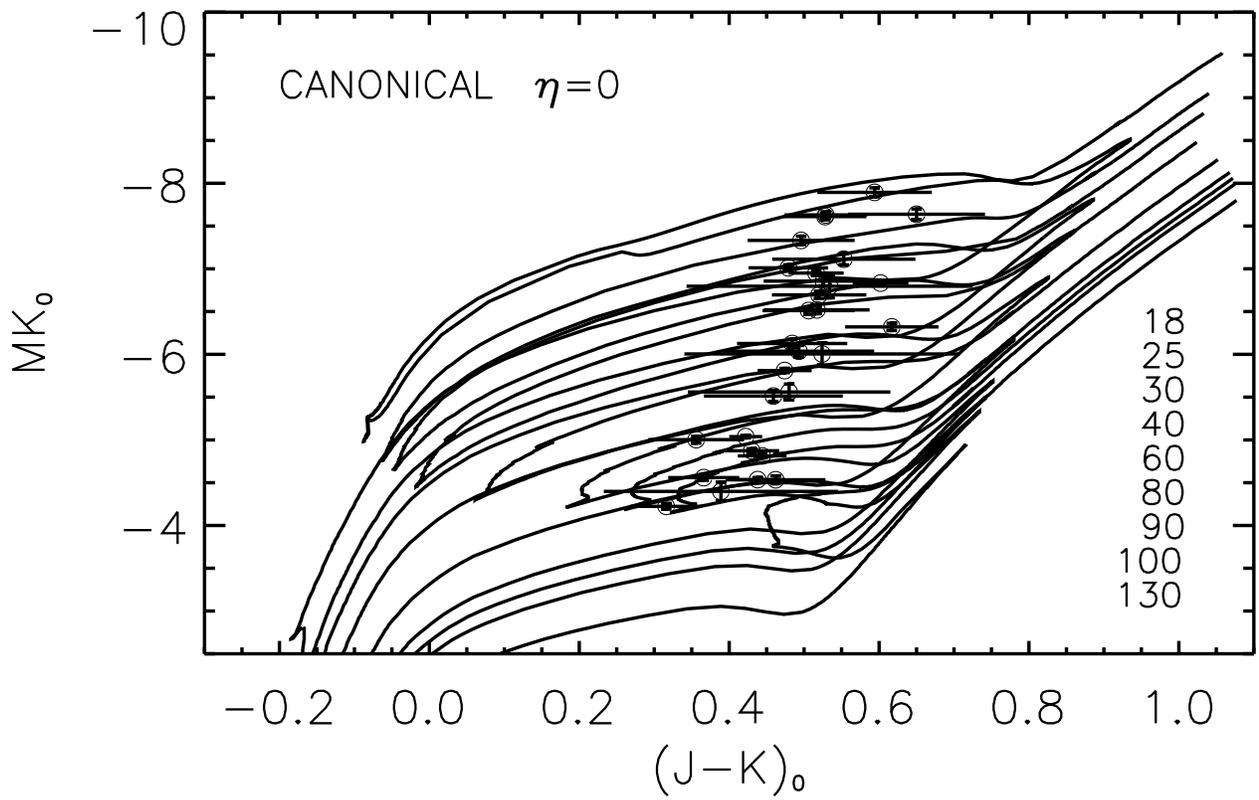
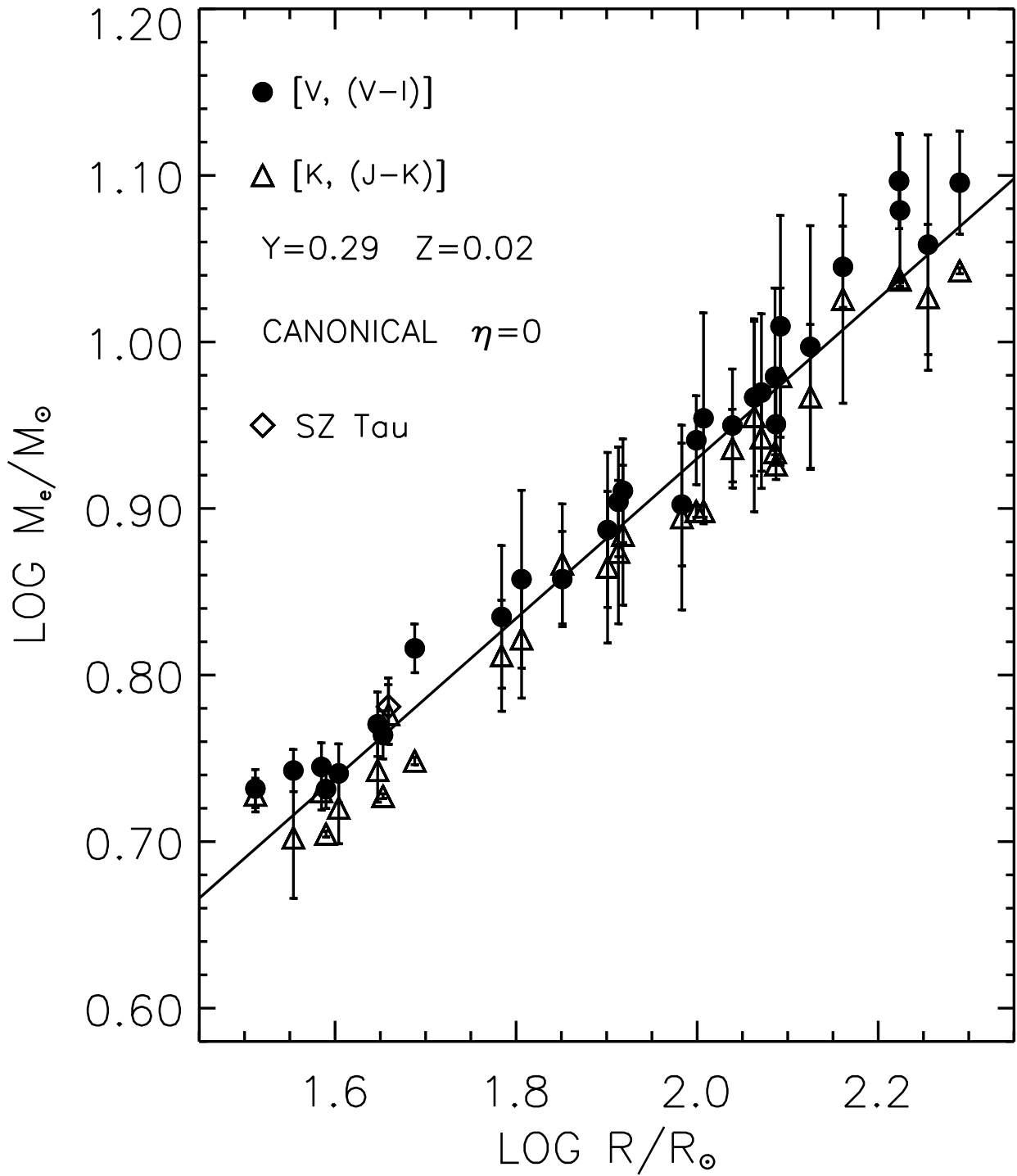
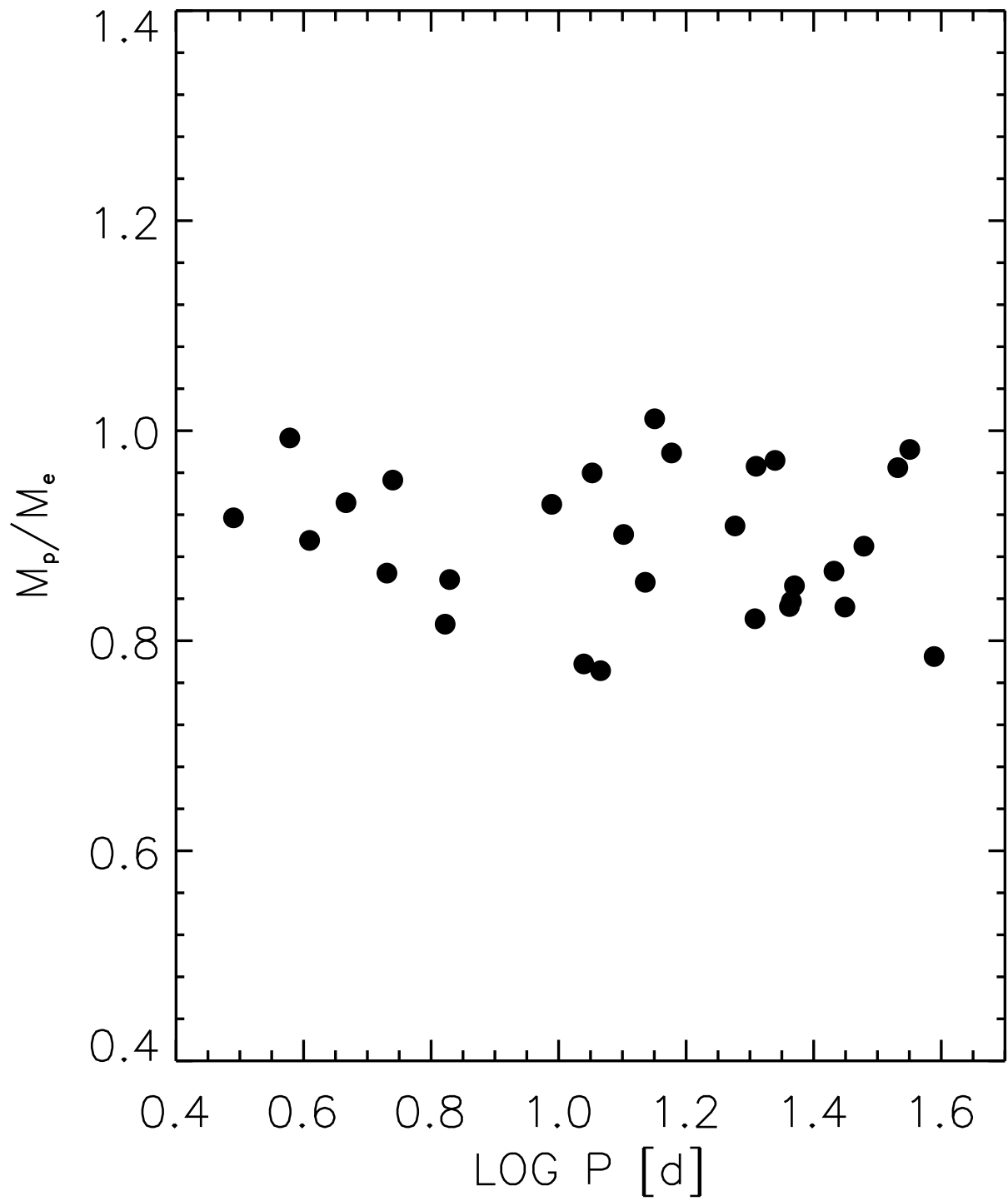
\end{document}